# On the optimization of hyperparameters in Gaussian process regression with the help of low-order high-dimensional model representation

Sergei Manzhos[1] and Manabu Ihara

School of Materials and Chemical Technology, Tokyo Institute of Technology, Ookayama 2-12-1, Meguro-ku, Tokyo 152-8552 Japan

**Abstract**

When the data are sparse, optimization of hyperparameters of the kernel in Gaussian process regression by the commonly used maximum likelihood estimation (MLE) criterion often leads to overfitting. We show that choosing hyperparameters (in this case, kernel length parameter and regularization parameter) based on a criterion of the completeness of the basis in the corresponding linear regression problem is superior to MLE. We show that this is facilitated by the use of high-dimensional model representation (HDMR) whereby a low-order HDMR representation can provide reliable reference functions and large synthetic test data sets needed for basis parameter optimization even when the original data are few.

# 1    Introduction

The use of the Gaussian process regression (GPR)[1] method has been gaining more and more traction in recent years in diverse applications.[2–4] This includes many of the traditional applications of machine learning as well as applications where particularly high accuracy is required. In particular, certain applications in quantum and computational chemistry require accuracies on the order of 0.01% or better; examples are spectroscopically accurate potential energy surfaces (PES)[5–11] and kinetic energy functionals (KEF) for orbital-free density functional theory (OF-DFT).[12–17] Especially with the recent appearance of comparisons between the popular neural network (NN) approach and GPR, which highlighted GPR advantages, in particular, in obtaining highly accurate approximations of multidimensional functions from few data,[18] the GPR method has received a boost in these applications. GPR is easy to use, in particular, in high-dimensional spaces; being a non-parametric method, the increase in space dimensionality does not lead to a drastic increase in the number of (non-linear)

---

[1] Author to whom correspondence should be addressed. E-mail: E-mail: manzhos.s.aa@m.titech.ac.jp , Tel & Fax : +81-3-5734-3918

parameters as, for example, in the case of neural networks (NN).[19] One problem area, as will also be highlighted below, is the optimal choice of GPR hyperparameters, especially when data are sparse. We note that data are always sparse in sufficiently high-dimensional spaces.[20] While the problem of GP hyperparameter optimization has been addressed with various methods, they typically focus on the issue of optimization in a multidimensional hyperparameter space, in the presence of noise, and of handling the cost of the evaluation of the reward function. Examples are random search related methods,[21] various versions of Bayesian inference[22,23] (with the commonly used maximum likelihood estimator, also used in the following, belonging to this type of methods[24]), simulated annealing,[25] genetic algorithms,[26] so-called bandit-based methods,[27] and combinations thereof.[28] Here, we address the issue of hyperparameter optimization from a different angle, namely, that whatever the choice of the reward (or penalty) function and whatever the cost of its evaluation, the sparsity of available training (and test) data may not permit optimizing the kernel parameters for the best *global* (i.e. in all relevant descriptors space) quality of function representation which can be considered from the point of view of the completeness of the basis set formed by the kernel functions (see Eq. 7 below), *even when the space of hyperparameters is low-dimensional* (only two-dimensional in the problem considered below) and when the data are not noisy.

We begin by briefly summarizing the basics of GPR. The problem we consider is building a faithful approximation of a continuous function $f(x), x \in R^D$ from a finite number of known samples $f^{(j)} = f(x^{(j)})$ at points $x^{(j)}, j = 1, \ldots, M$. In GPR, the expectation values $f(x)$ and variances $\Delta f(x)$ of the function values at any point in space $x$ are computed as[1]

$$f(x) = K^* K^{-1} f$$

(1)

$$\Delta f(x) = K^{**} - K^* K^{-1} K^{*T}$$

(2)

where $f$ is a vector of all (known) $f^{(j)}$ values, the matrix $K$ and row vector $K^*$ are computed from pairwise covariances among the data:

$$K = \begin{pmatrix} k(x^{(1)}, x^{(1)}) + \delta & k(x^{(1)}, x^{(2)}) & \cdots & k(x^{(1)}, x^{(M)}) \\ k(x^{(2)}, x^{(1)}) & k(x^{(2)}, x^{(2)}) + \delta & & k(x^{(2)}, x^{(M)}) \\ \vdots & & \ddots & \vdots \\ k(x^{(M)}, x^{(1)}) & k(x^{(M)}, x^{(2)}) & \cdots & k(x^{(M)}, x^{(M)}) + \delta \end{pmatrix}$$

$$\boldsymbol{K}^* = \begin{pmatrix} k(\boldsymbol{x}, \boldsymbol{x}^{(1)}) & k(\boldsymbol{x}, \boldsymbol{x}^{(2)}) & ... & k(\boldsymbol{x}, \boldsymbol{x}^{(M)}) \end{pmatrix},$$

(3)

(4)

and $K^{**} = k(\boldsymbol{x}, \boldsymbol{x})$. The covariance function $k(\boldsymbol{x}^{(1)}, \boldsymbol{x}^{(2)}|\boldsymbol{\lambda})$ is the kernel of GPR that depends on hyperparameters $\boldsymbol{\lambda}$ (which we omitted in the formulas for notational simplicity). The optional $\delta$ on the diagonal has the meaning of the magnitude of Gaussian noise and is a regularization (hyper)parameter; it helps generalization. Commonly used kernels belong to the Matern family:

$$k(\boldsymbol{x}, \boldsymbol{x}') = \sigma^2 \frac{2^{1-\nu}}{\Gamma(\nu)} \left( \sqrt{2\nu} \frac{|\boldsymbol{x} - \boldsymbol{x}'|}{l} \right)^{\nu} K_{\nu} \left( \sqrt{2\nu} \frac{|\boldsymbol{x} - \boldsymbol{x}'|}{l} \right)$$

(5)

where $\Gamma$ is the gamma function, and $K_{\nu}$ is the modified Bessel function of the second kind. At different values of $\nu$, this function becomes a squared exponential ($\nu \to \infty$), a simple exponential ($\nu = 1/2$) and various other widely used kernels (such as Matern3/2 and Matern5/2 for $\nu = 3/2$ and $5/2$, respectively). The value of $\nu$ is often preset, and the length scale $l$ and prefactor $\sigma^2$ are hyperparameters (i.e. $\boldsymbol{\lambda} = (l, \sigma^2)$) that can be optimized (as only relative values of $\delta$ and $\sigma^2$ are important, in principle, only one of them needs to be optimized, it will be $\delta$ in this work). Note that Eq. 2 as written (and as it usually appears in the literature) strictly speaking holds when $\sigma^2$ in Eq. 5 is the variance of the data; another way of reading Eq. 2 is to state that it is for normalized data (to unit variance).

When the hyperparameters are optimized, typically it is done by maximizing the log likelihood function (the so-called MLE estimator[24]):

$$\max \left( \frac{1}{2} ln|\boldsymbol{K}| - \frac{1}{2} \boldsymbol{t} \boldsymbol{K}^{-1} \boldsymbol{t} - \frac{M}{2} \ln(2\pi) \right)$$

(6)

With the right choice of the hyperparameters, GPR is able to obtain a smaller global error (on a large test set) than NN or, conversely, achieve a similar error with fewer data.[18] The last point is significant, as data are always sparse in sufficiently high-dimensional spaces in any realistic setting due to the curse of dimensionality. "Always sparse" here means that simply adding more data does not fix the data density issue under any realistic scenario; in the example of a 15-dimensional space considered below, a set of 5,000 points has the density of less than 1.8 data per dimension while increasing the set size to 50,000 would only increase the density to less than 2.1 data per dimension (of an equivalent direct product grid). It is also significant, because the computational cost of GPR (both of training the model and of recall) increases rapidly with the number of training data $M$ (due to the need to compute and wield the matrix $\boldsymbol{K}^{-1}$ of size $M \times M$ and corresponding matrix-vector and vector products). The issue of low data density can be palliated with RS-HDMR (random sampling high-dimensional model representation – see below)[29,30] -based approaches: RS-HDMR-GPR[31,32] and HDMR-type kernels[33,34]; these approaches (which are not limited to GPR[35–38]) allow building accurate approximations from extremely sparse data (down to 2 data per dimension or less). What remains critical is the choice of hyperparameters. We show below that the approach of Eq. 6 fails to find optimal hyperparameters when data are sparse and results in significant overfitting. This is in principle expected, as Eq. 6 only considers the training data and is ignorant of the shape of the function in all relevant space. The failure, however, happens at data densities which are practically relevant in applications, in particular, in the field of PES construction which we consider here as an example of a regression problem requiring high accuracy. It also happens at data densities where manually choosing hyperparameters does allow obtaining highly accurate global representations without overfitting. The failure is therefore not that of the GPR approach but of the method of hyperparameter optimization. We propose an alternative view of hyperparameter optimization problem as a problem of basis completeness and show that the optimal hyperparameters (kernel length scale $l$ and regularization parameter $\delta$) which avoid overfitting can be rationalized in this way. We also show that a low-order HDMR model, which can be built from sparse data without overfitting and which is then computable at any number of points anywhere in space, can be used as a reference function to optimize basis completeness. In particular, it can be used to generate large synthetic test data sets which can be used for basis parameter optimization.

## 2   Methods

Below, we show that Eq. 6 fails to find appropriate hyperparameters. In fact, the results we obtained with Eq. 6 are completely inacceptable. The difficulty of hyperparameter optimization with Eq. 6 is its reliance on the training data only; Eq. 6 is ignorant of overfitting by construction. Using a test or a validation set might allow identifying overfitting but would not address the issue in substance, as in high-dimensional spaces any test data would be sparse. We show below that this is not an issue of setting the right value of $\delta$ or optimizing it. We also show that the global quality of the approximation is best with the parameters which can be rationalized from the perspective of the quality of a basis set. Indeed, GPR is equivalent to a regularized linear regression. Eq. 1 can be written in the form of a basis expansion,

$$f(x) = \sum_{n=1}^{M} b_n(x) c_n$$

(7)

with basis functions $b_n(x) = k(x, x^{(n)})$ and with linear coefficients $c$ obtained with least squares, $c = K^{-1}f$.[39] The quality of the approximation is fully determined by the extent of the completeness of the basis set $\{k(x, x^{(n)}|\lambda)\}$. Hyperparameter optimization should therefore strive to improve the completeness of the basis in all relevant space rather than optimize a quantity only dependent on a low-density training point set. In the case of a squared exponential kernel, for example, it will be the completeness of a Gaussian basis set with basis functions located at each of the $M$ training points and with shapes defined by the width parameter $l$.

The completeness of a basis can be qualified by considering the quality of representation of certain test functions in all space. For example, in the field of electronic structure calculations, Chong introduced a completeness profile

$$Y = \sum_m \left( \int g(x) \kappa_m(x) dx \right)^2$$

(8)

to probe the completeness of the basis $\{\kappa_m\}$ when representing a GTO (Gaussian-type orbital) functions $g$.[40] Here the integral is over the support of the integrand. One wants $Y = 1$, and an incomplete basis will result in $Y < 1$. Conversely, one can look at this from the perspective of minimizing the error $e$ of representation of $g$ in $\{\kappa_m\}$, $e = 1 - \sum_m (\int g(x) \kappa_m(x) dx)^2$ or any other convenient definition of such error (root mean square, mean absolute etc.). We take a similar route. We will consider the quality of representation with Eq. 7 of a function similar in shape to $f(x)$, but, contrary to $f(x)$, known anywhere in space and such that it can be constructed from the same samples of $f(x)$ without overfitting. This function will be used to generate a large (unlimited in size) synthetic dataset that can be used to optimize the hyperparameters for the best quality of representation in all relevant space. To this end we use a low-order RS-HDMR[29,30,41] expansion of $f(x)$:

$$f(x) \approx f_0 + \sum_{i=1}^{D} f_i(x_i) + \sum_{1 \leq i < j \leq D} f_{ij}(x_i, x_j) + \cdots + \sum_{\{i_1 i_2 \ldots i_d\} \in \{12 \ldots D\}} f_{i_1 i_2 \ldots i_d}(x_{i_1}, x_{i_2}, \ldots, x_{i_d})$$

(9)

This is an expansion over orders of coupling of the variables. Taken to $d = D$, it is exact; when $d < D$ it is approximate. When $d$ is sufficiently low, this approximation can avoid overfitting even with few data, as low-order terms are well-defined with fewer data.[35] It is convenient to include lower-order terms into $d$-dimensional terms:

$$f(\pmb{x}) \approx \sum_{\{i_1 i_2 \ldots i_d\} \in \{12 \ldots D\}} f_{i_1 i_2 \ldots i_d}(x_{i_1}, x_{i_2}, \ldots, x_{i_d})$$

(10)

This approximation can be obtained with component functions $f_{i_1 i_2 \ldots i_d}$ constructed with different methods including ML methods like NN or GPR, giving rise to RS-HDMR-NN,[42] RS-HDMR-GPR[32] or GPR with an HDMR-type kernel.[33,34] We use here the latter approach for its simplicity, with $d = 1$, i.e. we use a kernel

$$k(\pmb{x}, \pmb{x}') = \sum_{i=1}^{D} k_i\ (x_i, x'_i)$$

(11)

That results in an approximation $f(\pmb{x}) \approx \sum_{i=1}^{D} f_i(x_i)$ with $f_i(x_i) = \pmb{K}_i^* \pmb{c}'$, where $\pmb{K}_i^*$ is a row vector with elements $k_i\ \left(x_i, x_i^{(n)}\right)$. In particular, the values of the component functions at the training set are $\pmb{f}_i = \pmb{K}_i \pmb{c}'$ where the $(m,n)$ elements of the matrix $\pmb{K}_i$ are $k_i\ \left(x_i^{(m)}, x_i^{(n)}\right)$. Here $\pmb{c}' = \left(\sum_{i=1}^{D} \pmb{K}_i\ \right)^{-1} \pmb{f}$. See Ref. [34] for details. We use this model as a reference function to optimize hyperparameters for basis completeness, that is we define

$$f_{ref}(\pmb{x}) = \left(\sum_{i=1}^{D} \pmb{K}_i^*\right) \pmb{c}'$$

(12)

As in most applications, the importance of terms of Eq. 9 rapidly drops with $d$,[31,32,35,43] hyperparameters (basis parameters) allowing for a good representation of $f_{ref}(\pmb{x})$ are expected to be good also for $f(\pmb{x})$.

The calculations were performed in Matlab 2021a using the *fitrgp* function. We chose as an example the fitting of the PES of $UF_6$ molecule to ab initio data. The PES is a 15-dimensional function, which we consider in the space of normal mode coordinates. The PES was sampled with about 50,000 ab initio calculations. The function values range 0 - 6,629 cm$^{-1}$. The details of the calculations, sampling point distribution etc. are given in Ref. [44]; they are not important for the purpose of this article except for the fact that the sampling was done using a quasirandom Sobol sequence,[45] i.e. the data are distributed in all space and not confirmed to any sub-dimensional hypersurfaces (it is important to note this in the context of the use of the RS-HDMR model in which all terms of Eq. 9 are obtained from one and the same data set). The data set is available in the Supporting Information of Ref. [31], where the plots describing the shape of the function can also be found. This is a convenient example because we previously reported the quality of fits using GPR, RS-HDMR-GPR, and GPR with HDMR-type kernel for these data, including approximate manual determination of the hyperparameters which provide the best *test* set error.[31,32,34] Those results can therefore serve as a reference and comparison point for the results we obtain here with basis completeness optimization. We used square exponential kernels in all models with a unit prefactor, $k(x, x') = exp(-|x - x'|^2/2exp(l)^2)$ and similarly for each $k_i(x_i, x'_i)$, i.e. the only kernel hyperparameter is the length parameter $l$; $\delta$ was also optimized in certain cases. The features are normalized to unit standard deviation, we therefore use isotropic kernels (same $l$ in all coordinates). Even though we have more than 50,000 data points available, we imagine that up to only 10,000 are available to tune the hyperparameters. This corresponds to a data density of less than 1.8 data per dimension (of an equivalent direct product grid). Some calculations use only 500 training points. 40,000 points are used to monitor the global quality of the approximation, i.e. we use a test set much larger than the training set and much larger than a smaller secondary test set (called below "train-test") used to tune the hyperparameters. We note that the ab initio data sampling the underlying function are not noisy, and that the number of hyperparameters is small. The difficulties of hyperparameter optimization demonstrated below are largely due to the low data density.

## 3 Results

In

Table 1, we list the results of GPR fits of the target function $f(x)$ and the reference functions $f_{ref}(x)$ with manually preset (scanned) hyperparameters ($l$ and $\delta$) and hyperparameters optimized with Eq. 6. In Table 2, we list the results of manual hyperparameter scanning and of using Eq. 6 when fitting a 1$d$ HDMR model $f_{ref}(x)$. In Figure 1 and Figure 2, we show correlation plots and Pearson correlation coefficients $R$ for selected cases. Note that there is some variability in the numbers due to the random

selection of train and test points. The extent of this variability is reported in Ref. [34] for the same data. The following follows from these results:

- There are hyperparameters ($l$ of about 3.0-3.5 and $\delta$ of about $1 \times 10^{-4...-5}$) which provide a global rmse (root mean square error) of below 40 cm$^{-1}$. The test set rmse and the ratio of train and test errors noticeably change as a function of $M$.

- MLE optimization fails to find reasonable values of hyperparameters. In fact, the global accuracy (test set rmse) when using Eq. 6 is so bad that the model is unusable, even with 10,000 training points (the results were naturally even worse with fewer training points). In other words, Eq. 6 cannot build a good basis even as it achieves a very low train set rmse, resulting in severe overfitting. It consistently underestimates $l$ and overestimates $\delta$. This is so even as we initialized the hyperparameters with values known to result in a good model (e.g. $l = 3.5$, $\delta = 1 \times 10^{-5}$). This is true and for the full-dimensional function $f(x)$, and for the additive $1d$ HDMR model $f_{ref}(x)$, even though the latter needs fewer data to be well determined. It is only with 10,000 training points and with fixed $\delta$ that MLE is able to find a good $l$ for the $1d$ HMDR model even though 500 points are quite enough to determine the 1$^{st}$ order HDMR component functions. One can view this from the positive side: the HDMR model allowed reasonable hyperparameter optimization via MLE while in a full-dimensional GPR, MLE resulted in sever overfitting even with 10,000 points.

- The reference $1d$ RS-HDMR model provides practically no overfitting, with train and test rmse within 10% of each other. The error of this model, of about 235 cm$^{-1}$, and the ratio of train and test errors display stability with respect to the number of training points $M$, confirming the thesis that fewer data are sufficient to determine low-order component function (compared to the full-dimensional function $f(x)$). As few as 500 points are sufficient to build an accurate additive model (with accuracy limited by HDMR order $d$ and not by the density of training data).

Out of the 10,000 data that we consider available for hyperparameter (basis) optimization, we reserve 5,000 as a test set which would have been available to tune the hyperparameters based on $f_{ref}(x)$. We call this a "test-train" set. We find that with 5,000 training points, the best test-train rmse of an additive model (Eq. 12) of about 230 cm$^{-1}$ is achieved with $l = 3 - 4$ and $\delta = 1 \times 10^{-4}$ (Table 2). The test rmse (on 40,000 points) is then also about 230 cm$^{-1}$ and is also optimal. We then fit a full-dimensional GPR model to the $f_{ref}(x)$ (which is now known anywhere in space) computed, using $l = 3$ and $\delta = 1 \times 10^{-4}$, at the 40,000 test points and manually optimize (scan) the hyperparameters.

Table 1 lists the results for $M$ = 5,000; they are similar for $M$ = 10,000. Naturally, the quality of the fit is very high (both in terms of rmse shown in the table and in terms of correlation between target and prediction shown in Figure 1), as the reference function is smooth. The train set rmse can be gotten very close to 0, while the test set rmse each $l$ levels off at as $\delta$ is decreased. We select a point where the test rmse is lowest while not exceeding the train rmse by more than two times (cf. the entries for the full dimensional fit of $f(x)$ with $M$ = 10,000, $l$ = 3.0-3.5, $\delta = 1\times10^{-4...-5}$). This happens at $l$ = 4.5, $\delta = 1\times10^{-6}$ where the test rmse is 1.7 cm$^{-1}$ and the train rmse is 1.2 cm$^{-1}$; calculations with $\delta = 1\times10^{-7}$ are unstable due to a high condition number of $K$ (high extent of overlap between basis functions centered at different points).

Table 1. Train and test set rmse values, in cm$^{-1}$, with different choices of hyperparameters, when fitting a full-dimensional GPR model to the target function $f(x)$ and the reference function $f_{ref}(x)$. The number of test points is 40,000 in all cases. Where the hyperparameters were optimized with Eq. 6, it is so indicated, and the optimized values are given. Optimal choices of hyperparameters for each $M$ are highlighted in bold. The results with hyperparameters based on $f_{ref}(x)$ are underlined.

| | | | rmse, cm$^{-1}$ | |
|---|---|---|---|---|
| $M$ | $l$ | $\log\delta$ | Train | Test |
| | | Target: $f(x)$ | | |
| 5,000 | 2.5 | -2 / -3 / -4 / -5 | 56.1 / 13.3 / 3.3 / 1.4 | 73.6 / 40.2 / 45.0 / 129.0 |
| 5,000 | **3.0** | -2 / -3 / **-4** / -5 | 148.2 / 41.7 / **10.1** / 3.2 | 167.1 / 60.3 / **36.9** / 51.9 |
| 5,000 | **3.5** | -3 / -4 / **-5** / -6 | 81.5 / 37.7 / **8.9** / 3.4 | 96.3 / 55.1 / **38.1** / 53.3 |
| 5,000 | 4.0 | -3 / -4 / -5 / -6 | 198.1 / 53.2 / 31.5 / 8.0 | 212.6 / 66.7 / 52.1 / 40.1 |
| 5,000 | <u>4.5</u> | -3 / -4 / -5 / <u>-6</u> | 231.4 / 136.6 / 49.4 / <u>26.7</u> | 239.0 / 155.3 / 63.3 / <u>47.8</u> |
| 10,000 | 2.5 | -2 / -3 / -4 / -5 | 53.4 / 14.6 / 4.5 / 2.2 | 61.2 / 27.0 / 28.9 / 55.4 |
| 10,000 | **3.0** | -2 / -3 / **-4** / -5 | 120.7 / 40.9 / **13.7** / 4.8 | 130.3 / 51.2 / **25.5** / 28.3 |
| 10,000 | **3.5** | -3 / -4 / **-5** / -6 | 65.5 / 35.0 / **13.5** / 5.5 | 73.1 / 45.3 / **25.8** / 27.9 |
| 10,000 | 4.0 | -3 / -4 / -5 / -6 | 180.5 / 52.7 / 29.4 / 13.5 | 185.7 / 60.9 / 40.5 / 26.4 |
| 10,000 | <u>4.5</u> | -3 / -4 / -5 / <u>-6</u> | 229.9 / 109.4 / 52.4 / 24.6 | 234.8 / 117.9 / 59.5 / <u>35.1</u> |
| 10,000 | ≥1.0 | -7 | unstable | |
| 10,000 | 0.5 | -7 | 1.1 | 812.7 |
| 5,000 | -0.79 (Eq. 6) | -5 | 0.1 | 1255.5 |
| 5,000 | 0.68 (Eq. 6) | 3.13 (Eq. 6) | 1337.7 | 1352.3 |
| 10,000 | 2.40 (Eq. 6) | -5 | 0.0 | 1261.7 |
| 10,000 | 0.69 (Eq. 6) | 3.13 (Eq. 6) | 1344.1 | 1350.9 |
| | | Target: $f_{ref}(x)$ | | |

| 5,000 | 3.0 | -3 / -4 / -5 | 7.5 / 1.2 / 0.1 | 9.6 / 3.7 / 0.87 |
| 5,000 | 3.5 | -3 / -4 / -5 | 12.6 / 2.6 / 0.5 | 14.9 / 3.6 /1.9 |
| 5,000 | 4.0 | -4 / -5 / -6* | 5.2 / 1.3 / 0.3 | 22.5 / 6.2 /2.1 / 1.6 |
| 5,000 | **4.5** | -4 / -5 / **-6*** | 18.3 / 1.9 / **1.2** | 19.8 / 2.4 / **1.7** |
| 5,000 | 5.0 | -4 / -5 / -6* | 22.1 / 11.2 / 1.6 | 23.0 / 12.9 / 2.0 |

*calculations with log $\delta$ = -7 are unstable

We also underline in Table 1 the results of GPR fits of the original target function $f(x)$ using the optimal hyperparameters determined via the additive HDMR model $f_{ref}(x)$. We observe that the hyperparameters that resulted in an optimal basis of Eq. 7 for $f_{ref}(x)$ also result in a good, albeit slightly suboptimal, basis for $f(x)$, and are much superior to the results of MLE optimization. This demonstrates that indeed a low-order HDMR model, which can be reliably constructed (i.e. without overfitting) with few data, can be used to optimize hyperparameters for basis completes for the GPR of the target function $f(x)$.

Table 2. Train and test set rmse values, in cm$^{-1}$, with different choices of hyperparameters, when fitting a $1d$ HDMR GPR model $f_{ref}(x)$ to the target function $f(x)$. The number of test points is 40,000 in all cases. Where the hyperparameters were optimized with Eq. 6, it is so indicated, and the optimized values are given. Optimal choices of hyperparameters are highlighted in bold.

| | | | rmse, cm$^{-1}$ | | |
| --- | --- | --- | --- | --- | --- |
| $M$ | $l$ | $\log\delta$ | Train | Test | Test-train* |
| 500 | 2.5 | -2 / -3 / -4 | 247.4/210.6/201.5 | 256.6/245.9/251.0 | 250.7/242.1/250.0 |
| 500 | 3.0 | -2 / -3 /-4 | 532.2/228.7/233.5 | 543.6/238.6/248.6 | 539.7/238.3/243.4 |
| 500 | 3.5 | -3 / -4 / -5 | 234.3/220.9/226.5 | 253.2/244.2/248.1 | 256.0/245.9/243.8 |
| 500 | 4.0 | -3 / -4 / -5 | 378.2/216.1/227.7 | 389.6/240.4/244.1 | 395.5/239.0/246.8 |
| 1,000 | 3.0 | -3 / -4 / -5 | 247.0/222.7/225.6 | 241.1/242.2/241.6 | 239.5/239.6/239.3 |
| 5,000 | **3.0** | -3 / **-4** / -5 | 239.7/231.5/230.2 | 236.6/235.3/236.0 | 234.5/**232.7**/239.7 |
| 5,000 | 3.5 | -3 / -4 / -5 | 233.1/227.8/230.6 | 234.9/235.1/235.4 | 233.2/234.0/235.8 |
| 5,000 | **4.0** | -3 / **-4** / -5 | 232.1/234.8/228.5 | 238.2/234.0/234.7 | 234.6/**229.9**/242.0 |
| 5,000 | 1.07 (Eq. 6) | -5.00 | 9.2 | 1320.7 | |
| 5,000 | 2.48 (Eq. 6) | 3.13 (Eq. 6) | 1345.3 | 1353.9 | |
| 10,000 | 2.94 (Eq. 6) | -5.00 | 229.3 | 235.7 | |
| 10,000 | 4.52 (Eq. 6) | 3.13 (Eq. 6) | 1360.9 | 1348.6 | |

*a test set assuming that only 5,000 test points are available

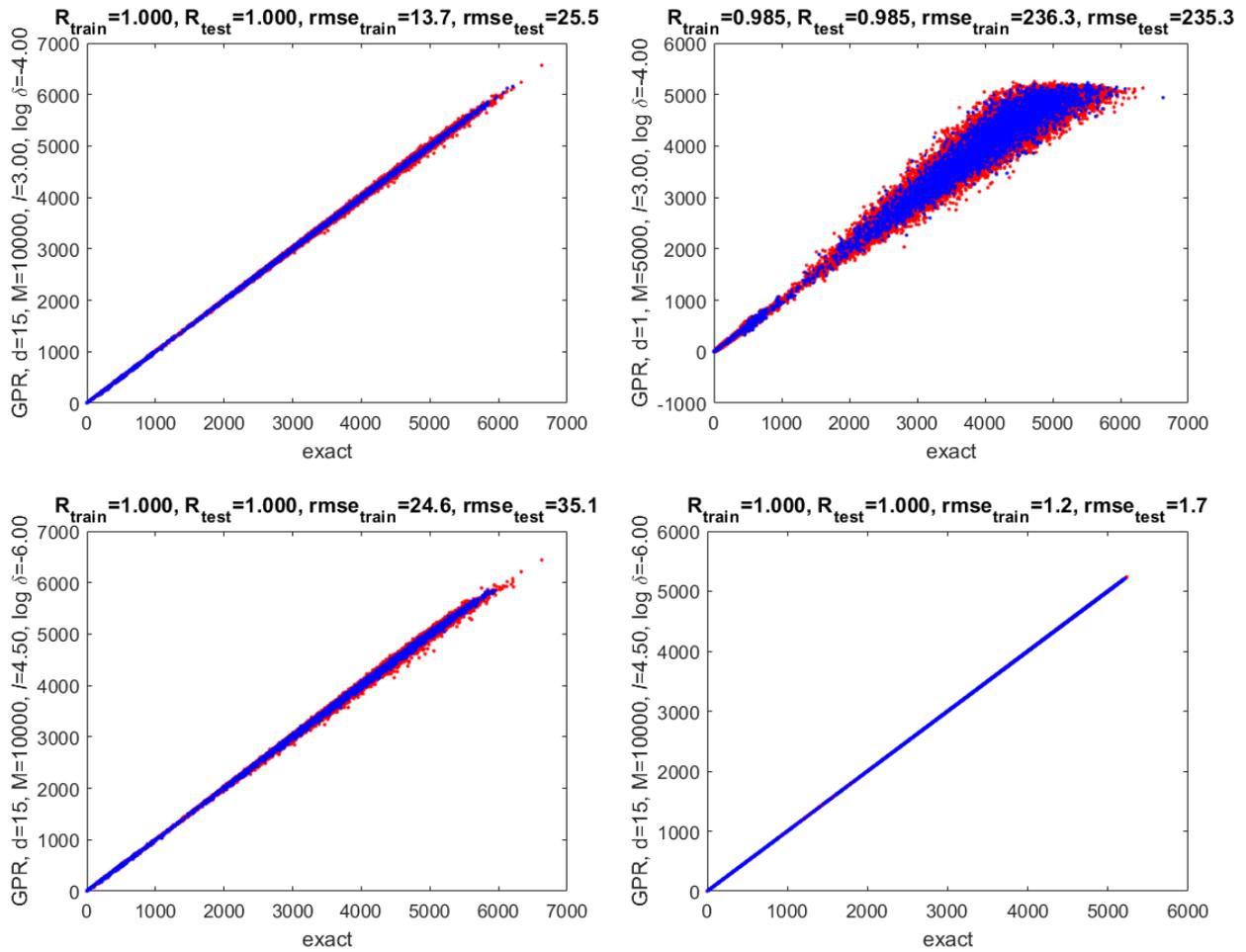

Figure 1. Example correlation plots between exact and predicted values of the target function with the full-dimensional GPR ("$d = 15$") and the HDMR model with $d = 1$ (Eq. 12, serving as a reference function $f_{ref}(x)$), with manually optimized hyperparameters (top panels) and with hyperparameters optimized for $f_{ref}(x)$ (bottom panels). The bottom left panel is the fit of $f(x)$ and bottom right – of $f_{ref}(x)$. Training points are shown in blue color and test points in red.

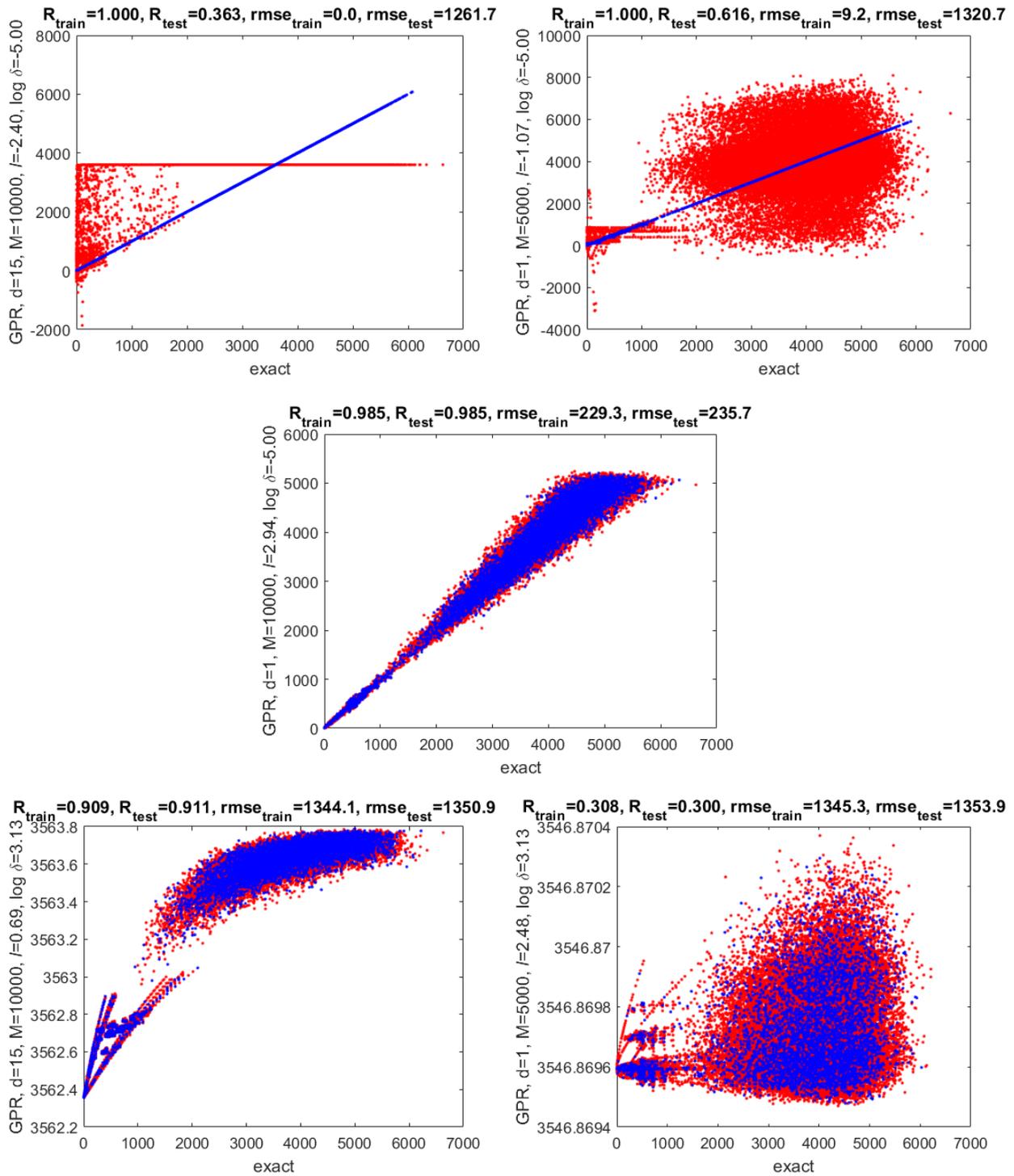

Figure 2. Example correlation plots between exact and predicted values of the target function with the full-dimensional GPR ("$d = 15$") and the HDMR model with $d = 1$ (Eq. 12), when the length parameter $l$ is optimized with Eq. 6 for fixed $\delta$ (top 3 panels of which the panel in the middle shows the only successful MLE optimization, with the 1-$d$ HDMR and with $M = 10,000$ and fixed $\delta$) and when both the length parameter and $\delta$ are optimized with Eq. 6 (bottom panel). Training points are shown in blue color and test points in red.

## 4  Conclusions

We have shown that when building a function with Gaussian process regression from sparse data in multidimensional spaces, the often-used MLE optimization of hyperparameters fails to build a good model. We have shown that a better approach is to optimize the hyperparameters from the perspective of the basis set made of the covariance functions. In this approach, a reference function or functions are needed whose values are accessible in all relevant space (i.e. allowing generation of very large test sets). We have shown that a low-order RS-HDMR model can be used as such a reference function, allowing finding hyperparameters which are also good for the representation of the target function. In the considered example of fitting a fifteen-dimensional interatomic potential of $UF_6$ with only 5,000 data (less than 1.8 data points per degree of freedom), we showed that this approach allows obtaining a similarly small global error as hyperparameter scanning which could have been done with the knowledge of a much larger test set, when using a reference function based on the 1$^{st}$ order HDMR model. The hyperparameters which are optimal for the 1$^{st}$ order HDMR model are sub-optimal but "good enough" for the target function. The approach is extendable to higher-order HDMR reference functions (depending on data availability) which would allow to optimize the basis even better.

What to do when the density of sampling even with many test points is still too low? In the case considered here, even the full set of about 50,000 points in a 15-dimensional space is quite sparse (less than 2.1 data points per degree of freedom). Of course, as the test function ($f_{ref}(\boldsymbol{x})$) can be computed anywhere, the test point set can be expanded at will. In practice, however, this may not be necessary. Our results also indicate that a good basis is that with the largest length parameters which can be used before numeric instability sets in. This result is in line with what we observed using Gaussian-type and other bell-shaped basis functions with the collocation method to solve the Schrödinger equation.[18,46,47] Another way to rationalize this is to consider that the expressive power of the basis is the highest when basis functions have non-negligible amplitudes at most data points. The parameter $\delta$ is then used to ensure the stability of the inverse of $\boldsymbol{K}$ for such a basis. This criterion could also be used as a rule of thumb to select the hyperparameters of a GPR kernel.

## 5  Acknowledgements

We thank Dr. Laura Laverdure and Prof. Nicholas Mosey who computed the data for Ref. [44] used here. Prof. Tucker Carrington and M. Eita Sasaki are thanked for discussions.